\documentstyle[amsfonts,preprint,aps]{revtex}

\begin{document}
\title{Realization of a General Multi-step Quantum Cloning Machine}
\author{L.\ Masullo, M. Ricci and \ F. De Martini}
\address{Dipartimento di Fisica and Istituto Nazionale per la Fisica della\\
Materia, Universit\`{a} di Roma ''La Sapienza'', p.le A. Moro 5, Roma\\
I-00185, Italy}
\maketitle

\begin{abstract}
A general multi-step $N\longmapsto M$ \ probabilistic optimal universal
cloning protocol is presented together with the experimental realization of
the \ $(1\rightarrow 3)$ and $(2\rightarrow 3)$ machines. Since the present
method exploits the bosonic nature of the photons, it can be applied to any
particle obeying to the Bose statistics. On a technological perspective, the
present protocol is expected to find applications as a novel, multi-qubit
symmetrizator device to be used in modern quantum information networks.
\end{abstract}


A most relevant limitation in quantum information processing is the
impossibility of perfectly cloning (copying) any unknown qubit $\left| \phi
\right\rangle $ \cite{Woot82}. Even if this process is unrealizable in its
exact form, it can be approximated {\it optimally} by the so-called
universal optimal quantum cloning machine $(UOQCM)$, one which exhibits the
minimum possible noise for any\ possible input state. From a theoretical
perspective, two different kinds of universal cloning machines have been
developed so far: a deterministic $N\rightarrow M$ UOQCM based on a unitary
operator acting on $N$ input qubits and $2(M-N)$ ancilla qubits \cite
{Buze96,Gisi97} and a {\it probabilistic} UOQCM based on a symmetrization
procedure involving a projective operator acting on $N$ inputs and $(M-N)$
blank ancilla qubits \cite{Wern98}.

In the last years several experimental realizations of the UOQCM for
polarization $(\pi -)$ encoded photon qubits have been reported. The
deterministic UOQCM has been realized by associating the cloning effect with
QED stimulated emission \cite{Lama02} while the probabilistic machine has
been realized adopting a linear symmetrization protocol \cite{Ricc04}. Thus
far, only the simplest $1\rightarrow 2$ cloning processes, i.e. for $N=1$
and $M=2$, were realized by both schemes. In particular, the probabilistic
process was achieved by exploiting the bosonic character of the photons
within a linear Hong-Ou-Mandel interferometer scheme \cite{HOM}.

The present work presents the first generalization of the universal optimal
cloning process by the realization of a very general linear procedure to
extend the probabilistic protocol to {\it any value} of $N$ and $M$
according to a suggestion by Werner. \cite{Wern98}. The validity of this
theoretical scheme is supported by the here reported experimental
implementations of the $1\rightarrow 3$ and $2\rightarrow 3$ probabilistic
processes for $\pi -$encoded photon qubits ($\pi -$qubits).

Let us outline first the $N\rightarrow M$ probabilistic cloning theory.
Consider $N$ identically prepared unknown qubits in the state $\rho
_{i}=\left| \phi \right\rangle \left\langle \phi \right| $ as input of the
cloning machine while $(M-N)$ blank qubits, i.e. all in the state $\rho _{A}=%
\frac{{\Bbb I}}{2}$, are used as an auxiliary resource. To generate $M$
output clones the machine performs the symmetrization of the output state by
applying the projective operator, $\Pi _{+}^{(M)}$ over the symmetric
subspace of $M$ qubits: 
\begin{equation}
\left| \phi \right\rangle \left\langle \phi \right| ^{\otimes N}\stackrel{%
_{N\rightarrow M}}{\longrightarrow }\frac{1}{p_{N\rightarrow M}}[\Pi
_{+}^{(M)}(\left| \phi \right\rangle \left\langle \phi \right| ^{\otimes
N}\otimes \frac{{\Bbb I}}{2}^{\otimes (M-N)})\Pi _{+}^{(M)}]  \label{eq1}
\end{equation}

where $p_{N\rightarrow M}=\frac{1}{2^{M-N}}\frac{1+M}{1+N}$ is the success
probability of the procedure. All the identical output clones are described
by the same output density matrix $\sigma ^{N\rightarrow M}={\cal F}%
_{N\rightarrow M}\left| \phi \right\rangle \left\langle \phi \right| +(1-%
{\cal F}_{N\rightarrow M})\left| \phi ^{\perp }\right\rangle \left\langle
\phi ^{\perp }\right| $, where ${\cal F}_{N\rightarrow M}$= $\langle \phi
|\sigma ^{N\rightarrow M}\left| \phi \right\rangle $ = $(N+1+\beta )/(N+2)$
with $\beta \equiv N/M\leq 1\;$is the ''fidelity'' of the optimal cloning
process \cite{Buze96,Gisi97,Brub98}. In the case of our present experiment,
the value of these parameters for the $1\rightarrow 3$ and $2\rightarrow 3$
machines are found respectively: $p_{1\rightarrow 3}=\frac{1}{2},$ ${\cal F}%
_{1\rightarrow 3}=\frac{7}{9},\ p_{2\rightarrow 3}=\frac{2}{3},$ ${\cal F}%
_{2\rightarrow 3}=\frac{11}{12}$.

In order to implement the process expressed by Eq.\ref{eq1}, consider that
any generic{\it \ }$1\rightarrow M${\it \ }cloning process can in facts be
realized by a chain of {\it \ }$\left( M-1\right) ${\it \ }intermediate {\it %
identical} machines according to the operatorial identity: $\Pi _{+}^{(M)}$=$%
\Pi _{+}^{(M)}(\Pi _{+}^{(M-1)}\otimes {\Bbb I}^{(1)})$=$\Pi _{+}^{(M)}(\Pi
_{+}^{(M-1)}\otimes {\Bbb I}^{(1)})(\Pi _{+}^{(M-2)}\otimes {\Bbb I}%
^{(2)})\cdot \cdot (\Pi _{+}^{(2)}\otimes {\Bbb I}^{(M-2)})$. This is
justified by the very definition of \ the symmetric subspace of $M${\it \ }%
qubits as the smallest subspace in{\it \ }$H_{d}^{\otimes M}${\it \ }spanned
by the tensor product vectors $\left| \phi \right\rangle ^{\otimes M}${\it \ 
}for {\it any }$\left| \phi \right\rangle \in H_{2}$, the qubit space with
d=2.{\it \ }In the above expression ${\Bbb I}^{(i)}${\it \ }stands for the
identity operator in the{\it \ }$i-$qubit space,{\it \ }$H_{2}^{\otimes 
\text{ }i}$.{\it \ }This implies that the symmetrization of $M$\ qubits can
be carried out step by step e.g. starting from the symmetrization of \ the
two input qubits $\rho $ and $\overline{\rho }$, as shown in Fig. 1a \cite
{Brub98}.{\it \ }Precisely, the state $\varrho ^{(i)}$ realized{\it \ }at
the output of any $i^{th}$ machine in the chain, i.e. of the overall $%
1\rightarrow i$ device, belongs to the set: 
\begin{eqnarray}
\varrho ^{(2)} &=&\Pi _{+}^{\left( 2\right) }(\rho \otimes \overline{\rho }%
)\Pi _{+}^{\left( 2\right) },..\text{ }\varrho ^{(i)}=\Pi _{+}^{(i)}(\varrho
^{(i-1)}\otimes \overline{\rho })\Pi _{+}^{(i)},..\varrho ^{(M)}=\Pi
_{+}^{(M)}(\varrho ^{(M-1)}\otimes \overline{\rho })\Pi _{+}^{(M)}= 
\nonumber \\
&&\Pi _{+}^{(M)}(\Pi _{+}^{(M-1)}\otimes {\Bbb I}^{(1)})\cdot \cdot (\Pi
_{+}^{(2)}\otimes {\Bbb I}^{(M-2)})[\rho \otimes \overline{\rho }^{\otimes
(M-1)}](\Pi _{+}^{(2)}\otimes {\Bbb I}^{(M-2)})\cdot \cdot (\Pi
_{+}^{(M-1)}\otimes {\Bbb I}^{(1)})\Pi _{+}^{(M)}
\end{eqnarray}
Note that in the above expressions the input states, i.e. the {\it pure} $%
\rho \equiv \left| \phi \right\rangle \left\langle \phi \right| $ and the 
{\it fully mixed} $\overline{\rho }\equiv \frac{{\Bbb I}}{2}$, can be
interchanged leading to the two equivalent configurations shown in Fig.1a:
the upper one has been chosen for the present implementations. These schemes
are represented by arrays of equal Hong-Ou-Mandel interferometers, each one
consisting of a 50/50 beam-splitter $(BS)$ and realizing the qubit
symmetrization in Hilbert spaces of increasing dimensions. The theory above
can be easily extended to the analysis of any general $N\rightarrow M$
cloning process. Optionally, the procedure could be made to consist of a
sequence of linear symmetrization devices acting by {\it inequal} cloning
steps, e.g. by injection of different mixed states along the chain.

The Fig. 1b shows the experimental apparatus by which the 2-step chain $%
(1\rightarrow 3)=(1\rightarrow 2)+(2\rightarrow 3)${\it \ }UOQCM\ has been
realized.

$1\rightarrow 2$ $UOQCM$. The device realizing the first step
state-symmetrization was the beam-splitter $BS_{A}$ excited over the two
input modes by the 2-qubit state: $\rho _{in}^{1\rightarrow 2}=$ $\left|
\phi \right\rangle \left\langle \phi \right| _{S}\otimes \frac{{\Bbb I}_{A}}{%
2}$. After projection by $BS_{A\text{ }}$in the symmetric subspace, the
output state realized on the output mode $k_{2}$ was expressed by: 
\begin{equation}
\rho _{out}^{1\rightarrow 2}\equiv \varrho ^{(2)}=\frac{2}{3}\left| \phi
\phi \right\rangle \left\langle \phi \phi \right| +\frac{1}{3}\left| \{\phi
\phi ^{\perp }\}\right\rangle \left\langle \{\phi \phi ^{\perp }\}\right|
\label{eq3}
\end{equation}
where the notation $\left| \left\{ \phi \phi ^{\perp }\right\} \right\rangle 
$ stands for a total symmetric combination of the states $\left| \phi
\right\rangle $ and $\left| \phi ^{\perp }\right\rangle $. The identical
condition realized on mode $k_{1}$ was neglected, for simplicity. The two
clones $j=1,2$ emitted over $k_{2}$ were expressed by the same operators: $%
\sigma _{j}^{1\rightarrow 2}$=$Tr_{h\neq j}\rho _{out}^{1\rightarrow 2}=%
\frac{5}{6}\left| \phi \right\rangle \left\langle \phi \right| +\frac{1}{6}%
\left| \phi ^{\perp }\right\rangle \left\langle \phi ^{\perp }\right| $.

In the experiment\ a pair of photons was generated by spontaneous parametric
down conversion (SPDC) in a 1 mm thick BBO crystal, cut for Type I phase
matching. The two photons, each with wavelength (wl) $\lambda =795nm$ and
coherence time $\tau _{coh}\simeq 200fs$, were emitted over two modes $k_{A}$
and $k_{S}$ respectively in the {\it product state} of horizontal $(H)$
linear polarizations $(\pi )$: $\left| H\right\rangle _{S}\left|
H\right\rangle _{A}$. Then, on mode $k_{S}$ the qubit $\left| H\right\rangle
_{S}$ was $\pi -$encoded by an optical waveplate (wp) $WP_{\phi }$ into the
generic {\it pure-state} $\left| \phi \right\rangle _{S}$, $\rho _{S}=\left|
\phi \right\rangle \left\langle \phi \right| _{S}$, while on mode $k_{A}$
the qubit $\left| H\right\rangle _{A}$ was transformed into a fully {\it %
mixed-state} $\rho _{A}=\frac{{\Bbb I}_{A}}{2}$ by a depolarizing channel
realized by a stochastically driven Electro-Optics Pockels $(EOP)$\ cell, $%
P_{A}$. The two qubits $\rho _{S}$ and $\rho _{A}$ were then drawn into a
linear superposition in $BS_{A}$. The exact space-time overlap of the two
input modes implying the actual realization of the interference was
controlled by the microscopic $BS_{A}$ displacement: $Z_{A}=2c\Delta t$.
Let's call ''$BS_{A}$-{\it interference}'' the condition $Z_{A}=0$
corresponding to maximum interference. By turning on the cloning machine,
i.e. setting it in $BS_{A}$-{\it interference}, the induced {\it Bose
coalescence} implied an enhancement by a factor $R_{1\rightarrow 2}=2$ of
the $\left| \phi \phi \right\rangle $ component in the 2-qubit output state
and no enhancement of the $\left| \{\phi \phi ^{\perp }\}\right\rangle $
component \cite{Ricc04}. The measurement of $R_{1\rightarrow 2}$ was carried
out by a post-selection technique, by the $\pi -$analysis setup shown at the
r.h.s. of Fig.1b, connected directly to the output mode $k_{2}$, by
disregarding at this stage the presence of $BS_{B}$. This $\pi -$analyzer
consisted of an output mode selector, realized by a 5 meter long single-mode
optic fiber, followed by the wp $WP_{\phi }^{-1}$ that mapped the output
state $\left| \phi \right\rangle $ into $\left| H\right\rangle $ by
counterbalancing the action of the input $WP_{\phi }$. Finally, by a {\it %
polarizing-}$BS$, $(PBS)$\ the $\left| H\right\rangle $ and $\left|
V\right\rangle $ components of the output state were directed respectively
over the modes $k_{\phi }$ and $k_{\phi }^{\ast }$. The mode $k_{\phi }$ was
coupled to the\ detectors $D_{1}$, $D_{2}$, $D_{3}$ by means of two equal
50/50 beam-splitters $BS_{1}$, $BS_{2}$ while the mode $k_{\phi }^{\ast }$
was coupled to $D_{1}^{\ast },D_{2}^{\ast }$ by $BS_{3}.$ The $\left| \phi
\phi \right\rangle $ component was identified by detecting coincidence by
the $D-$pair sets $[D_{1},D_{2}]$ and $[D_{1},D_{3}]$ while the state $%
\left| \{\phi \phi ^{\perp }\}\right\rangle $ was identified by $%
[D_{1},D_{1}^{\ast }]$. The detectors $(D)\;$were equal single-photon
counters SPCM-AQR14.

Three different input states $\left| \phi \right\rangle _{S}=\left|
H\right\rangle $, $2^{-%
{\frac12}%
}(\left| H\right\rangle +\left| V\right\rangle )$, $2^{-%
{\frac12}%
}(\left| H\right\rangle +i\left| V\right\rangle ),$ identified in the
following by{\Large \ }$\left| H\right\rangle ,\left| H+V\right\rangle $%
{\Large \ }and{\Large \ }$\left| H+iV\right\rangle ${\Large \ }respectively,%
{\Large \ }were adopted to test the {\it universality} of the device. The
cloning process was found to affect only the $\left| \phi \phi \right\rangle 
$ component, as expected, and $R_{1\rightarrow 2}\;$was determined as the
ratio between the peak value (cloning machine switched on) and the basis
value (off). The corresponding experimental values of the {\it cloning
fidelity, }$F_{1\rightarrow 2}=\left( 2R_{1\rightarrow 2}+1\right) /\left(
2R_{1\rightarrow 2}+2\right) $ were: $F_{1\rightarrow 2}^{H}=0.831\pm 0.001$;%
$\ F_{1\rightarrow 2}^{H+V}=0.833\pm 0.002$; $F_{1\rightarrow
2}^{H+iV}=0.830\pm 0.002$. These values are to be compared with the
theoretical value $F_{1\rightarrow 2}^{th}=5/6\approx 0.833$ corresponding
to the optimal enhancement ratio $R=2$. Similar results for the $%
1\rightarrow 2$ UOQCM have been reported in \cite{Ricc04}.

$1\rightarrow 3$ $UOQCM$. In agreement with the upper configuration shown in
Fig.1a, the 50/50 beam-splitter $BS_{B}$ was the next state-symmetrization
device: the whole $(1\rightarrow 3)\ UOQCM$ is shown in Fig.1b. This $BS$
was excited over the input mode $k_{2}$ by the output state $\rho
_{out}^{1\rightarrow 2}$ of \ the$1\rightarrow 2$ UOQCM, Eq. \ref{eq3}, and
over the other input $k_{B}$ by the fully {\it mixed state} $\rho _{B}=\frac{%
{\Bbb I}_{B}}{2}$. In analogy with the first step experiment, this state was
obtained by means of a $EOP$, $P_{B}$ acting on a highly attenuated quasi
single-photon beam expressed by the $\pi -$qubit $\left| H\right\rangle _{B}$%
, deflected from the main laser by the mirror $M$, and delayed by $%
Z_{B}=2c\Delta t_{B}$ via an ``optical trombone''. Once again, the condition 
$Z_{B}=0$, dubbed here as ''$BS_{B}$-{\it interference'' }condition{\it , }%
was made to correspond to the maximum overlapping of the 2 input modes of $%
BS_{B}$. In particular, in {\it no}-''$BS_{A}$-{\it interference'' }%
condition, i.e. for $\left| Z_{A}\right| \gg 2c\tau _{coh}$, the ''$BS_{B}$-%
{\it interference'' }corresponded{\it \ } to the maximum overlapping in $%
BS_{B}$ of the $mixed$ states $\rho _{A}$ and $\rho _{B}$. In summary, the
overall $1\rightarrow 3$ machine was excited by the input state $\rho
_{in}^{1\rightarrow 3}$= $\left| \phi \right\rangle \left\langle \phi
\right| _{S}\otimes \frac{{\Bbb I}_{A}}{2}\otimes $ $\frac{{\Bbb I}_{B}}{2}$%
, i.e. by the pure state $\rho _{S}$ to be cloned and by two mutually {\it %
uncorrelated} mixed states $\rho _{A}$ and $\rho _{B}$. By applying to this
state the projector $\Pi _{+}^{(3)}=\left| \phi \phi \phi \right\rangle
\left\langle \phi \phi \phi \right| +\left| \{\phi \phi \phi ^{\perp
}\}\right\rangle \left\langle \{\phi \phi \phi ^{\perp }\}\right| +\left|
\{\phi \phi ^{\perp }\phi ^{\perp }\}\right\rangle \left\langle \{\phi \phi
^{\perp }\phi ^{\perp }\}\right| +\left| \phi ^{\perp }\phi ^{\perp }\phi
^{\perp }\right\rangle \left\langle \phi ^{\perp }\phi ^{\perp }\phi ^{\perp
}\right| $, the symmetrized output state is obtained:

\begin{equation}
\rho _{out}^{1\rightarrow 3}\equiv \varrho ^{(3)}=\frac{3}{6}\left| \phi
\phi \phi \right\rangle \left\langle \phi \phi \phi \right| +\frac{2}{6}%
\left| \{\phi \phi \phi ^{\perp }\}\right\rangle \left\langle \{\phi \phi
\phi ^{\perp }\}\right| +\frac{1}{6}\left| \{\phi \phi ^{\perp }\phi ^{\perp
}\}\right\rangle \left\langle \{\phi \phi ^{\perp }\phi ^{\perp }\}\right|
\label{eq4}
\end{equation}
Each one of the identical clones $j=1,2,3$ can be thought to be expressed by
the reduced density matrix: $\sigma _{j}^{1\rightarrow 3}$=$Tr_{h,k\neq
j}\rho _{out}^{1\rightarrow 3}$=$\frac{7}{9}\left| \phi \right\rangle
\left\langle \phi \right| $ + $\frac{2}{9}\left| \phi ^{\perp }\right\rangle
\left\langle \phi ^{\perp }\right| $. The projection over the symmetric
subspace was identified by the measurement of the 3-photon Fock state over
the output mode $k_{3}$ by the $\pi -$analyzer apparatus already described
and shown at the r.h.s. of Fig.1b. The output field emitted over $k_{4}$ was
negleced. The $\left| \phi \phi \phi \right\rangle $, $\left| \{\phi \phi
\phi ^{\perp }\}\right\rangle $ and $\left| \{\phi \phi ^{\perp }\phi
^{\perp }\}\right\rangle $ components of $\rho _{out}^{1\rightarrow 3}$ were
measured by the 3-fold coincidence events respectively by the detector $(D)$%
\ sets $[D_{1},D_{2},D_{3}]$, $[D_{l},D_{m},D_{n}^{\ast }]$ and $%
[D_{l},D_{n}^{\ast },D_{p}^{\ast }]$ for any $l,m=1,2,3$ and $n,p=1,2$. A
little inspection of the circuit leads to the following expectations. In the
exact $BS_{B}${\it -resonance, }and{\it \ no-}$BS_{A}${\it -resonance, }i.e.
in the condition ''Bose coalescence'' of only the two {\it mixed} states $%
\rho _{A}$and $\rho _{B}$ in $BS_{B}$, an enhancement by a factor $\Gamma $
of \ the $\left| \phi \phi \phi \right\rangle $ and $\left| \{\phi \phi
^{\perp }\phi ^{\perp }\}\right\rangle $ components\ should be detected by
the $\pi -$analyzer. Furthermore, by turning on also the $BS_{A}${\it %
-resonance}, i.e. by setting $Z_{A}$= $Z_{B}$= $0$, a further enhancement of
the $\left| \phi \phi \phi \right\rangle $ component by a factor $%
R_{1\rightarrow 3}^{1}$and of the $\left| \{\phi \phi \phi ^{\perp
}\}\right\rangle $ component by a factor $R_{1\rightarrow 3}^{2}$ were
expected. In summary, the full resonance condition, corresponding to the
swithing on of the $\Pi _{+}^{(3)}$ projector, implied the global
enhancements by the factors $\Gamma R_{1\rightarrow 3}^{1},${\large \ }$%
R_{1\rightarrow 3}^{2}${\large \ }and{\large \ }$\Gamma $ respectively of
the components $\left| \phi \phi \phi \right\rangle ${\large , }$\left|
\{\phi \phi \phi ^{\perp }\}\right\rangle ${\large \ }and{\large \ }$\left|
\{\phi \phi ^{\perp }\phi ^{\perp }\}\right\rangle ${\large \ }of the state $%
\rho _{out}^{1\rightarrow 3}$, Eq.\ref{eq4}. Accordingly, the first step of
our strategy consisted of the measurement of $\Gamma $. This was provided by
injecting in the apparatus the {\it pure} state $\left| \Psi \right\rangle
_{SAB}\equiv $ $\left| V\right\rangle _{S}\left| H\right\rangle _{A}\left|
V\right\rangle _{B}$, by setting $Z_{A}$= $Z_{B}$= $0$ and by turning off
the mixing $EOP\;$devices $P_{A}$ and $P_{B}$. The value of $\Gamma $ was
determined by the ratio of the counting rates of the $\left| \{\phi \phi
\phi ^{\perp }\}\right\rangle $ components, the only non vanishing one under 
$\left| \Psi \right\rangle _{SAB}$ excitation. These rates were measured at
the peak of the detected resonance curves, i.e. with $Z_{SB}=0$, and far
from the peak, with $Z_{SB}>>2c\tau _{coh}$, being $Z_{SB}=Z_{A}-Z_{B}$ \
the mutual delay between qubits $S$ and $B$ at $BS_{B}$. The measured value $%
\Gamma ^{\exp }=$1.66$\pm 0.05$, \ expressing the degree of
indistinguishability attained between photons coming form different sources,
SPDC and attenuated laser, was \ to be compared with the theoretical one $%
\Gamma ^{th}=2$.\ By restoring the full operation of the overall apparatus
under excitiation by $\rho _{in}^{1\rightarrow 3}$, $R_{1\rightarrow 3}^{1}$
and $R_{1\rightarrow 3}^{2}$ were determined as the ratios of the $\left|
\phi \phi \phi \right\rangle $ and $\left| \{\phi \phi \phi ^{\perp
}\}\right\rangle $ component measured, via $3$-$D$ coincidences, in
resonance, i.e. $Z_{B}=Z_{A}=0$ and out of resonance, i.e. $Z_{B}=0,$ $%
Z_{A}>>2c\tau _{coh}$. From these measurements, the fidelity of the overall
process could be determined: $F_{1\rightarrow 3}$ = $(3\Gamma
R_{1\rightarrow 3}^{1}+4R_{1\rightarrow 3}^{2}+\Gamma )/(3\Gamma
R_{1\rightarrow 3}^{1}+6R_{1\rightarrow 3}^{2}+3\Gamma )$. The plots of Fig.
2a show the experimental $3$-$D$ coincidence results measured by the $\pi -$%
analyzer for the various state components of the output state: $\rho
_{out}^{1\rightarrow 3}$. The experimental values of the fidelity measured
by the above procedure under excitation of three different input states $%
\rho _{S}=\left| \phi \right\rangle \left\langle \phi \right| _{S}$ were: $%
F_{1\rightarrow 3}^{H}=0.758\pm 0.008$;$\ F_{1\rightarrow 3}^{H+V}=0.761\pm
0.003$; $F_{1\rightarrow 3}^{H+iV}=0.758\pm 0.008$ to be compared with the
optimum value: $F_{1\rightarrow 3}^{th}=7/9\approx 0.778$.

The protocol can be easy generalized for any long $UOQCM$\ chain by a
straightforward repetition of the above procedure, as follows. The output
state $\rho _{out}^{1\rightarrow 3}$ of the $1\rightarrow 3$ $UOQCM$ be
injected into one input arm of a further state-symmetrizing 50/50
beam-splitter $BS_{C}$ while a mixed one-photon state $\rho _{C}=\frac{{\Bbb %
I}_{C}}{2}$ be injected on the other arm. This state is generated, as
previously, by extracting by a further mirror $M$ a highly attenuated beam
expressed by the $\pi -$qubit $\left| H\right\rangle _{C}$ and mixing it by
an $EOP$, $P_{C}$. This will result in a $1\rightarrow 4$ $UOQCM$ apparatus
generating the output state $\rho _{out}^{1\rightarrow 4}$ .Then again: the
output state $\rho _{out}^{1\rightarrow 4}$of the $1\rightarrow 4$ $UOQCM$
be injected into one input arm of a state-symmetrizing \bigskip $BS_{D}$
while a mixed state $\rho _{D}=\frac{{\Bbb I}_{C}}{2}$....and so on.

$2\rightarrow 3$ $UOQCM$. As a significant variant of the above protocol,
the $P_{A}$ Pockels Cell was removed and the beam-splitter and $BS_{A}$ was
excited over the input modes $k_{A}$ and $k_{S}$ by the {\it same} pure
states: $\varrho _{A}=\varrho _{S}=\left| \phi \right\rangle \left\langle
\phi \right| $. In $BS_{A}-${\it resonance} condition, i.e. $Z_{A}=0$, the $%
BS_{A}$ acted as a conventional Hong-Ou-Mandel interferometer\ emitting over
the output mode $k_{2}$ the symmetric Bose state $\left| \phi \right\rangle
\left\langle \phi \right| _{2}^{\otimes 2}$ to be injected, together with $%
\rho _{B}=%
{\frac12}%
{\Bbb I}_{B}$ into the $BS_{B}$ according to the discussion above: Fig.1b.
The input state to this novel $N\rightarrow M$ $\ UOQCM$ with $N=2$ and $M=3$
was then expressible as: $\rho _{in}^{2\rightarrow 3}$= $\left| \phi
\right\rangle \left\langle \phi \right| _{2}^{\otimes 2}\otimes $ $\frac{%
{\Bbb I}_{B}}{2}$, to be mapped onto the 3 clone output state, according to:

\begin{equation}
\rho _{in}^{2\rightarrow 3}\rightarrow \rho _{out}^{2\rightarrow 3}=\frac{3}{%
4}\left| \phi \phi \phi \right\rangle \left\langle \phi \phi \phi \right| +%
\frac{1}{4}\left| \{\phi \phi \phi ^{\perp }\}\right\rangle \left\langle
\{\phi \phi \phi ^{\perp }\}\right|  \label{cloning23}
\end{equation}
Each output clone $j=1,2,3$ could be expressed by the state $\sigma
_{j}^{2\rightarrow 3}$=$Tr_{h,k\neq j}\rho _{out}^{2\rightarrow 3}$ = $\frac{%
11}{12}\left| \phi \right\rangle \left\langle \phi \right| +\frac{1}{12}%
\left| \phi ^{\perp }\right\rangle \left\langle \phi ^{\perp }\right| .$

Once again, the $\left| \phi \phi \phi \right\rangle $ and $\left| \{\phi
\phi \phi ^{\perp }\}\right\rangle $ components of $\rho
_{out}^{2\rightarrow 3}$ were measured by the $3$-$D$ coincidence events
respectively by the sets $[D_{1},D_{2},D_{3}]$ and $[D_{l},D_{m},D_{n}^{\ast
}]$ for any $l,m$=1,2,3 and $n=1,2$. The plots shown in Fig.2b express the
experimental results. In full analogy with the previous discussion, the
detected bosonic coalescence was expressed by an enhancement of the detected
coincidences by a factor $R_{2\rightarrow 3}=3$ of the $\left| \phi \phi
\phi \right\rangle $ component of $\rho _{in}^{2\rightarrow 3}$ whereas no
enhancement affected $\left| \{\phi \phi \phi ^{\perp }\}\right\rangle $.
Furthermore, in analogy with the $1\rightarrow 2$ cloning process, the
experimental value of $R_{2\rightarrow 3}\;$was determined as the ratio
between the peak resonant value for the $BS_{B}$ interferometer $(Z_{B}=0)$
and the out of resonance value: $Z_{B}>>2c\tau $. The universality condition
was assessed by injecting the same three different input test states adopted
for the previuos cases. The experimental values of fidelity were found : $%
F_{2\rightarrow 3}^{H}$= $0.895\pm 0.003$;$\ F_{2\rightarrow 3}^{H+V}$= $%
0.893\pm 0.003$; $F_{2\rightarrow 3}^{H+iV}$= $0.894\pm 0.003$ in
correspondence with the input states defined above. Tese values are be
compared with the calculated optimal value: $F_{2\rightarrow 3}^{th}$= $%
\left( 3R_{2\rightarrow 3}+2\right) /\left( 3R_{2\rightarrow 3}+3\right) $= $%
(11/12)\approx 0.917$. In all previous experiment, the measured values of $%
R_{2\rightarrow 3}$\ were reduced by the unwanted injection of two and three
photons in the mode $k_{B}$\ and by simultaneous emissions of two pairs from
SPDC. These spurious events affected the measured value of the $%
R_{2\rightarrow 3}$ factor by a calculated average amount of $\approx 15\%$.

In summary a very general and efficient linear multi-step optical procedure
for the probabilistic $N\rightarrow M$\ optimal universal cloning machine
has been proposed together with the successful experimental realization of
of the first two steps, i.e. the $(1\rightarrow 3)\;UOQCM$. Furthermore, the
probabilistic $(2\rightarrow 3)$\ $UOQCM$, was also demonstrated by a
straightforward variant of the same protocol. This shows that a very similar
protocol can be adopted to implement contextually the $N\rightarrow M$\
UOQCM, the $N\rightarrow (M-N)$\ Universal NOT gate and any {\it %
programmable anti-unitary map }\cite{Hill02} following the very general
symmetrization procedure recently proposed by \cite{Symm04}. Since the
present method basically exploits the bosonic nature of the photons, it can
be straightforwardly applied to any particle obeying to the Bose statistics.
On a more sophisticated technological perspective, the present protocol is
expected to find applications as a realization of a general, multi-qubit
device based on the state-symmetrization process to be used in modern
Quantum Information networks. \cite{Bare97,Cirac99,Puri04}.

{\bf Figure Captions:}

Fig.1.(a)Linear optical scheme for the realization of the general $%
1\rightarrow M$ and $M-1\rightarrow M$ Universal Quantum Cloning Machines by
a chain of {\it identical} symmetrizer beam splitters. (b) Experimental
set-up of a $1\rightarrow 3$ cloning process.

Fig.2. Experimental results of the $1\rightarrow 3$ and $2\rightarrow 3$
UOQCMs for three input qubits. (a) From the upper to the lower row: data
corresponding to $\left| \phi \phi \phi \right\rangle $, $\left| \{\phi \phi
\phi ^{\perp }\}\right\rangle $ and $\left| \{\phi \phi ^{\perp }\phi
^{\perp }\}\right\rangle $ components of $\rho _{out}^{1\rightarrow 3}$
measured by 3-fold coincidences. From the left to the right column: data
corresponding to the $\left| H\right\rangle $, $\left| H+V\right\rangle $, $%
\left| H+iV\right\rangle $ input state $\rho _{S}.${\Large \ }(b) From the
upper to the lower row: data corresponding to $\left| \phi \phi \phi
\right\rangle $\ and $\left| \{\phi \phi \phi ^{\perp }\}\right\rangle $\
components of $\rho _{out}^{2\rightarrow 3}$\ measured by 3-fold
coincidences. From the left to the right column: data corresponding to the $%
\left| H\right\rangle $, $\left| H+V\right\rangle $, $\left|
H+iV\right\rangle $\ input state $\rho _{S}.$

\bigskip

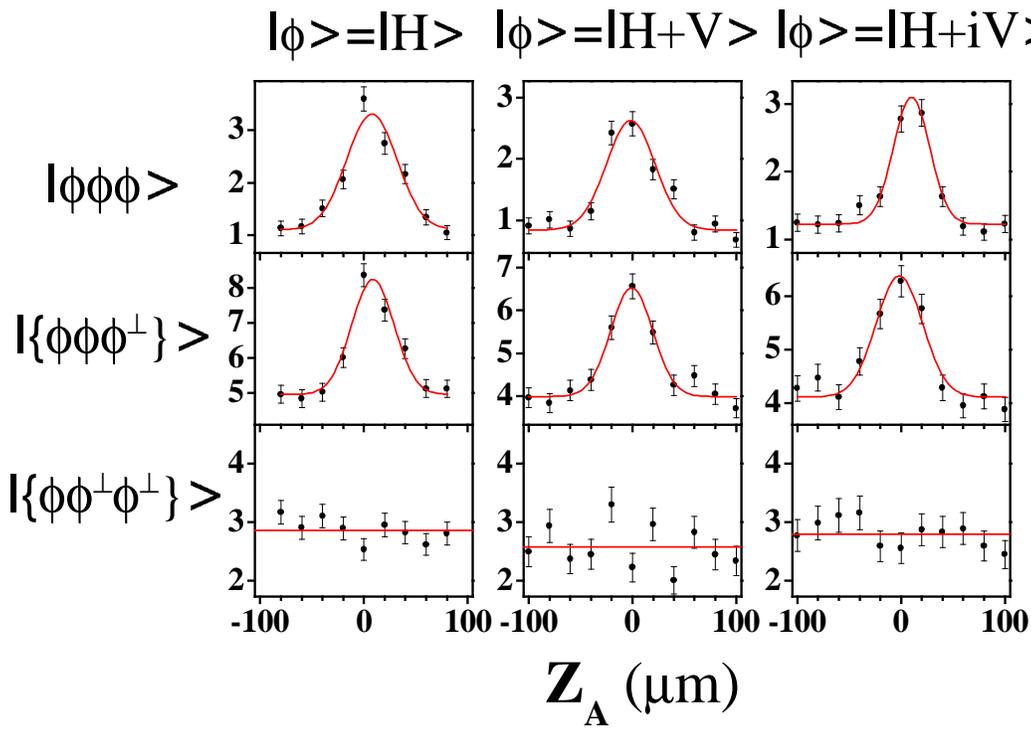
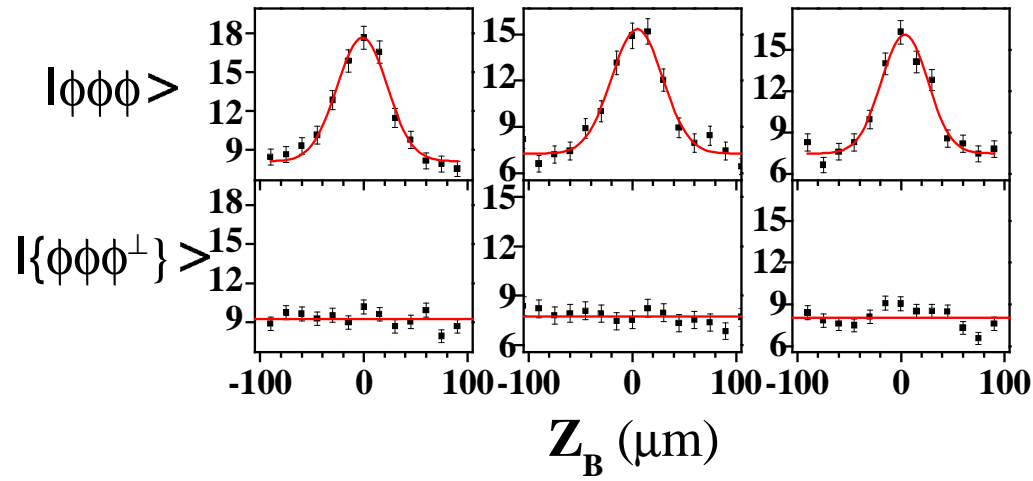
\end{document}